\documentclass[12pt,oneside,a4paper]{article}

\usepackage{hyperref}
\hypersetup{colorlinks,bookmarksopen,bookmarksnumbered,citecolor=blue,
linkcolor=black,pdfstartview=FitH,urlcolor=blue}

\usepackage[dvipdfmx]{graphicx}
\usepackage{amssymb}
\usepackage{cite}
\usepackage{bm}
\usepackage{indentfirst}
\usepackage{amsmath}
\usepackage{hhline}
\usepackage{multirow}
\usepackage{color}

\usepackage{braket}
\usepackage{textcomp}

\allowdisplaybreaks
\numberwithin{equation}{section}

\definecolor{dred}{rgb}{0.7,0.15,0.09}
\definecolor{dblue}{rgb}{0,0.12,0.64}
\definecolor{dgreen}{rgb}{0.2,0.51,0.19}

\begin{document}

\begin{titlepage}

\begin{flushright}
KANAZAWA-23-05\\
NITEP 169
\end{flushright}

\begin{center}

\vspace{1cm}
{\large\textbf{
Relation between higher-dimensional gauge theories and gravitational waves from first-order phase transitions
}
 }
\vspace{1cm}

\renewcommand{\thefootnote}{\fnsymbol{footnote}}
Takuya Hirose$^{1}$\footnote[1]{taku555phys@gmail.com}
and
Hiroto Shibuya$^{2}$\footnote[2]{h.shibuya.phys@gmail.com}
\vspace{5mm}

\textit{
 $^1${
 Nambu Yoichiro Institute of Theoretical and Experimental Physics (NITEP), \\
Osaka Metropolitan University, Osaka 558-8585, Japan}\\
 $^2${Institute for Theoretical Physics, Kanazawa University, Kanazawa 920-1192, Japan}
}

\vspace{8mm}

\abstract{
\noindent
In this work, we investigate the relation between higher-dimensional gauge theories and stochastic gravitational wave (GW) spectrums caused by their potential.
It is known that the higher-dimensional gauge theories can induce the spontaneous symmetry breaking of the gauge symmetry.
If the spontaneous symmetry breaking induces the first-order phase transition, the stochastic GW can be observed in future interferometers.
Through our numerical calculations, 
we reveal that distinctive parameters in the theories, like the compact scale, can change the GW spectrums dynamically. We also discuss the verifiability of the theories through the GW observations.
}

\end{center}
\end{titlepage}

\renewcommand{\thefootnote}{\arabic{footnote}}
\newcommand{\bhline}[1]{\noalign{\hrule height #1}}
\newcommand{\bvline}[1]{\vrule width #1}

\setcounter{footnote}{0}

\setcounter{page}{1}

\section{Introduction}
The Standard Model (SM) has succeeded in explaining phenomena related to particle physics.
However, the SM has some problems, hence some extension to the Beyond Standard Model (BSM) in the ultraviolet (UV) scale would be needed.
As a candidate for the BSM, higher-dimensional field theories have been studied \cite{Manton:1979kb, Arkani-Hamed:1998jmv, Randall:1999ee, Appelquist:2000nn, Hosotani:1983xw, Hosotani:1988bm, Hatanaka:1998yp}.
In the higher-dimensional theories, the spontaneous symmetry breaking (SSB) of the gauge symmetry occurs because of the dynamics of the Wilson line phases, called the Hosotani mechanism \cite{Hosotani:1983xw, Hosotani:1988bm}.
The higher-dimensional gauge theory (for example, gauge-Higgs unification, GHU \cite{Kubo:2001zc, Scrucca:2003ra, Haba:2004qf, Maru:2006wa, Hosotani:2007kn, Adachi:2018mby}) also can solve the gauge hierarchy problem.
Therefore, we can calculate the finite Higgs mass without invoking supersymmetry in the higher-dimensional gauge theory \cite{Hatanaka:1998yp, Maru:2006wa, Hosotani:2007kn}.
In the framework of higher-dimensional gauge theory at finite temperature, the scalar fields are possible to induce the first-order phase transitions in the SSB~\cite{Panico:2005ft, Maru:2005jy, Maru:2006wx, Adachi:2019apm, Funatsu:2021gnh}.

If the first-order phase transition occurs, it yields the stochastic gravitational wave (GW). The GW spectrums can be observed by the future space-based interferometers like the approved Laser Interferometer Space Antenna (LISA)~\cite{Caprini:2015zlo, LISA:2017pwj, Caprini:2019egz} and Deci-Hertz Interferometer Gravitational Wave Observatory (DECIGO)~\cite{Seto:2001qf, Kawamura:2011zz, Kawamura:2020pcg}, and the ground-based interferometer, Einstein telescope (ET)~\cite{Punturo:2010zz}. The GW peak frequency relates to a scale for the first-order phase transition.
Hence, the measurement for the wide GW frequency by combing observations would be important to find signatures of the BSM.

In this paper, we investigate relations between parameters in the higher-dimensional gauge theories and the GW spectrums from the first-order phase transitions.
The first-order phase transition occurs by a temperature change of a potential for the extra dimensional component of a gauge field.
Through our numerical calculations in simple setups,
we reveal that a four-dimensional gauge coupling in the higher-dimensional gauge theory can control the GW energy density.
Besides, we also confirm that a compact scale of the theory changes the GW frequency, while is unrelated to the shape of the GW spectrum as properties of GW spectrum from phase transitions.
Regions of the above parameters, which can be investigated in future interferometers, are also shown. It may help one to consider the verifiability of the higher-dimensional gauge theory through the GW observations.

The outline of this paper is as follows: 
We briefly review the five-dimensional theory in Sec.~\ref{sec:2} and give the thermal effective potential.
In Sec.~\ref{sec:3}, we introduce calculations for the GW spectrum from three sources when the SSB induces the first-order phase transition.
Sec.~\ref{sec:4} is devoted to showing numerical results. We reveal the relations between the distinctive parameters in the higher-dimensional gauge theories and the GW spectrums.
Lastly, we give a conclusion in Sec.~\ref{sec:5}

\section{Five-dimensional gauge theory} \label{sec:2}
In this section, we will review a five-dimensional $SU(3)$ gauge theory on $M^4\times S^1/Z_2$ \cite{Haba:2004qf},
where $M^4$ is a Minkowski spacetime and the orbifold $S^1/Z_2$ is a circle imposed $Z_2$ parity. 
The radius of the circle is written as $R$, which is called the compact scale.
We denote $x^\mu~(\mu=0,1,2,3)$ as the coordinates in Minkowski spacetime and $y$ as the fifth-dimensional coordinate.
Because of the $Z_2$ parity, two fixed points ($y=0$ and $y=\pi R$) appear.
Under the $Z_2$ transformation at $y=0$,
the bulk gauge fields $A_M(x^\mu,y)$, with the spacetime index $M=\mu,y$, transform as
	\begin{align}
	A_\mu(x^\mu,-y)&=P_0A_\mu(x^\mu,y)P^{-1}_0\,, \label{y=0gauge}\\
	A_y(x^\mu,-y)&=-P_0A_y(x^\mu,y)P^{-1}_0 \label{y=0WL}\,,
	\end{align}
where $P_0=P^{-1}_0=P^\dag_0$ denotes the operation of $Z_2$ transformation at $y=0$.
Under the $Z_2$ transformation at $y=\pi R$, the bulk gauge fields transform as
	\begin{align}
	A_\mu(x^\mu,\pi R-y)&=P_1A_\mu(x^\mu,\pi R+y)P^{\dag}_1\,, \label{y=piRgauge} \\
	A_y(x^\mu,\pi R-y)&=-P_1A_y(x^\mu,\pi R+y)P^{\dag}_1\,,\label{y=piRWL}
	\end{align}
where $P_1=P^{-1}_1=P^\dag_1$ is the operation of $Z_2$ transformation at $y=\pi R$.
On the other hand, we must impose the periodic boundary condition on the bulk gauge fields as
	\begin{align}
	A_M(x^\mu,y+2\pi R)=U A_M(x^\mu,y) U^\dag\,, \label{pbc}
	\end{align}
where $U$ is a unitary matrix.
Note that Eq.~\eqref{y=piRgauge} can be derived using Eqs.~\eqref{y=0gauge} and \eqref{pbc}, and then we can obtain the relation $U=P_1P_0$.

Due to the orbifold boundary conditions $P_0$ and $P_1$, the gauge symmetry is explicitly broken.
In this paper, we consider a $SU(3)$ gauge group with the orbifold boundary conditions:
	\begin{align}
	P_0=P_1=\text{diag}(1, 1, -1)\,.
	\end{align}
Denoting $(\pm,\pm)$ as the $Z_2$ charges at $y=0$ (left) and $y=\pi R$ (right),
$A_\mu$ and $A_y$ are decomposed as
	\begin{align}
	A_\mu&=\left(\begin{array}{cc|c}
	(+,+) & (+,+) & (-,-) \\
	(+,+) & (+,+) & (-,-) \\
	\hline(-,-) & (-,-) & (+,+)
	\end{array}\right)\,, \label{Amu.decom} \\
	A_y&=\left(\begin{array}{cc|c}
	(-,-) & (-,-) & (+,+) \\
	(-,-) & (-,-) & (+,+) \\
	\hline(+,+) & (+,+) & (-,-)
	\end{array}\right)\,. \label{Ay.decom}
	\end{align}
The bulk gauge field components $A_\mu$ and $A_y$ with $(+,+)$ parity have a four-dimensional massless zero mode. 
Hence we can read that the $SU(3)$ gauge symmetry is broken into $SU(2)\times U(1)$ from Eq.~\eqref{Amu.decom}.
While $A_y$ has a $(+,+)$ parity in the non-diagonal part (in terms of the $SU(3)$ generators, $\lambda^{4,5,6,7}$), this part is identified with a scalar doublet.

We suppose that the vacuum expectation value (VEV) of $A_y$ is taken as
\begin{align}
\braket{A_y}=\frac{a}{g_4R}\frac{\lambda^6}{2}, 
\label{VEV}
\end{align}
with a dimensionless real parameter $a$ and a four-dimensional coupling $g_4\equiv g/\sqrt{2\pi R}$. The coupling $g$ is a six-dimensional gauge coupling.
From the potential analysis in a later section, Eq.~\eqref{VEV} leads to the relation,
\begin{align}
\frac{a_0}{g_4R}= v,
\label{eq:relation}
\end{align}
where $a_0$ is a constant determined by the minimum of the potential, and $v$ is a field vacuum expectation value.
The parameter $a$ is related to the Wilson line phase $W$ and determines the pattern of gauge symmetry breaking.
The Wilson line is written as
	\begin{align}
	W=P\exp\left(ig\oint_{S^1}dy A_y\right)=\left(\begin{array}{ccc}
	1 & 0 & 0 \\
	0 & \cos(\pi a) & i\sin(\pi a) \\
	0 & i\sin(\pi a) & \cos(\pi a)
	\end{array}\right)\,.
	\end{align}
In the $a=0$ case, the Wilson line phase has a unit matrix, therefore the gauge symmetry $SU(2)\times U(1)$ remains unbroken.
In the $a=1$ case, the Wilson line phase has a matrix $\text{diag}(1,-1,-1)$ since $\cos(\pi a)=-1$ and $\sin(\pi a)=0$. 
It represents the SSB of the gauge symmetry, $SU(2)\times U(1)\rightarrow U'(1)\times U(1)$.
In the $a\ne1$ case, the gauge symmetry $SU(2)\times U(1)$ is broken into $U(1)$.
This symmetry breaking pattern can be applied to electroweak symmetry breaking $SU(2)_{L}\times U(1)_{Y}\rightarrow U(1)_{EM}$.

Using periodic boundary condition \eqref{pbc} and $Z_2$ transformation at $y=0$ and $y=\pi R$, we can expand $A_\mu$ in terms of Kaluza-Klein (KK) modes as
	\begin{align}
	A_{\mu}\left(x^{\mu}, y\right)_{(+,+)}&=\frac{1}{\sqrt{2\pi R}}A^{(0)}_\mu(x^\mu)_{(+,+)}+\frac{1}{\sqrt{\pi R}}\sum_{n=1}^\infty A^{(n)}_\mu(x^\mu)_{(+,+)}\cos\left(\frac{ny}{R}\right)\,, \label{Amu(++)expand} \\
	A_{\mu}\left(x^{\mu}, y\right)_{(-,-)}&=\frac{1}{\sqrt{\pi R}}\sum_{n=1}^\infty A^{(n)}_\mu(x^\mu)_{(-,-)}\sin\left(\frac{ny}{R}\right)\,. \label{Amu(--)expand}
	\end{align}
The expansion of the extra dimensional component $A_{y}\left(x^{\mu}, y\right)_{(+,+)}$ and $A_{y}\left(x^{\mu}, y\right)_{(-,-)}$ are the same expansion of Eqs.~\eqref{Amu(++)expand} and \eqref{Amu(--)expand}, respectively, since $Z_2$ assignments are same.
From the KK expansions and the VEV of $A_y$, the KK mass eigenvalues of $A_\mu$ are derived as
	\begin{align}
	\frac{n^2}{R^2}\times2,~\frac{(n\pm a )^2}{R^2},~\frac{(n\pm a /2)^2}{R^2}\times2\,. \label{gaugeKKmass}
	\end{align}
Based on the information of KK mass in Eq.~\eqref{gaugeKKmass}, we can calculate the four-dimensional effective potential of $A_y$ at zero temperature as~\cite{Haba:2004qf, Kubo:2001zc}
	\begin{align}
	V^{g}_\mathrm{eff}( a )
	&=-3C\sum_{n=1}^{\infty}\frac{1}{n^5}\Big(\cos(2\pi n a )+2\cos(\pi n a )\Big)\,,
	\end{align}
where $C=3/(64\pi^6R^4)$.

Considering extra matter fields,
their contributions to the effective potential can change the gauge symmetry breaking pattern.
Let us introduce the extra matter fields, which are $N_{f}$ fermions $\psi$ in the fundamental representation, $N_\mathrm{ad}$ fermions $\psi^a$ in the adjoint representation, and $N^{s}_{f}$ scalars $\phi$ in the fundamental representation.
The $Z_2$ transformations of these extra fields are given by
	\begin{align}
	&\phi(x,-y)=\eta P_0 \phi(x, y) \quad, \quad \phi(x, \pi R-y)=\eta^{\prime} P_1 \phi(x, \pi R+y)\,, \\
	&\psi(x,-y)=\eta P_0 \gamma^{5} \psi(x, y) \quad, \quad \psi(x, \pi R-y)=\eta^{\prime} P_1\gamma^{5} \psi(x, \pi R+y)\,, \\
	&\psi^{a}(x,-y)=\eta P_0 \gamma^{5} \psi^{a}(x, y) P^{\dagger}_0 \quad, \quad \psi^{a}(x, \pi R-y)=\eta^{\prime} P_1\gamma^{5} \psi^{a}(x, \pi R+y) P^{\dagger}_1\,,
	\end{align}
with $\eta,\eta'=\pm$.
The KK expansions of extra fields are different from the sign of $\eta\eta'$.
The fields with $\eta\eta'=+$ are the same expansion of Eqs.~\eqref{Amu(++)expand} and \eqref{Amu(--)expand}.
On the other hand, we can expand the fields with $\eta\eta'=-$ as
	\begin{align}
	\Phi\left(x^{\mu}, y\right)_{(+,-)}&=\frac{1}{\sqrt{\pi R}}\sum_{n=1}^\infty \Phi^{(n)}(x^\mu)_{(+,-)}\cos\left(\frac{\left(n+\frac{1}{2}\right)y}{R}\right)\,, \\
	\Phi\left(x^{\mu}, y\right)_{(-,+)}&=\frac{1}{\sqrt{\pi R}}\sum_{n=1}^\infty \Phi^{(n)}(x^\mu)_{(-,+)}\sin\left(\frac{\left(n+\frac{1}{2}\right)y}{R}\right)\,,
	\end{align}
where $\Phi$ can be replaced for $\phi$, $\psi$, and $\psi^a$.
As in the derivation of Eq.~\eqref{gaugeKKmass}, we can obtain the KK masses of the fields with $\eta\eta'=+$ or $\eta\eta'=-$ in the fundamental (adjoint) representation.
The effective potential of the contributions from matter fields is given by
	\begin{align}
	V^{m}_\mathrm{eff}( a )&=(4N^{(+)}_f-2N^{s(+)}_f)C\sum_{n=1}^{\infty}\frac{1}{n^5}\cos(\pi n a )
	+(4N^{(-)}_f-2N^{s(-)}_f)C\sum_{n=1}^{\infty}\frac{1}{n^5}\cos(\pi n( a -1)) \nonumber \\
	&\quad+4N^{(+)}_\mathrm{ad} C\sum_{n=1}^{\infty}\frac{1}{n^5}\Big(\cos(2\pi n a )+2\cos(\pi n a )\Big) \nonumber \\
	&\quad+4N^{(-)}_\mathrm{ad} C\sum_{n=1}^{\infty}\frac{1}{n^5}\Big(\cos(2\pi n( a -1/2))+2\cos(\pi n( a -1))\Big)\,.
	\end{align}
Here, $N_f$, $N_{\mathrm{ad}}$, and $N^s_f$ are decomposed into $N_f=N^{(+)}_f+N^{(-)}_f$, $N_{\mathrm{ad}}=N^{(+)}_{\mathrm{ad}}+N^{(-)}_{\mathrm{ad}}$, and $N^s_f=N^{s(+)}_f+N^{s(-)}_f$.
We define the total effective potential of $A_y$ at zero temperature as the combination of the contributions,
	\begin{align}
	V^{T=0}_\mathrm{eff}(a)\equiv V^g_\mathrm{eff}(a)+V^m_\mathrm{eff}(a).
	\end{align}

We derive the effective potential at finite temperature.
In the finite temperature field theories, the time direction is compactified on a circle $S^1$.
Through a radius of the circle $S^1$, we introduce a temperature $T$ and express the radius as $T^{-1}$.
Following Ref.\cite{Maru:2005jy, Antoniadis:2001cv}, the four-dimensional thermal effective potential can be obtained as
	\begin{align}
	V^{T\ne0}_\mathrm{eff}( a ,T)&=\frac{2\Gamma(5/2)}{\pi^{5/2}}\sum_{l=1}^{\infty}\sum_{n=1}^{\infty}\frac{1}{[(2\pi Rn)^2+(l/T)^2]^{5/2}} \nonumber \\
	&\quad\times
	\left[\Big(-3+4(-1)^l N^{(+)}_\mathrm{ad}\Big)\Big(\cos(2\pi n a )+2\cos(\pi n a )\Big)\right. \nonumber \\
	&\hspace{10mm}+4(-1)^l N^{(-)}_\mathrm{ad}\Big(\cos(2\pi n( a -1/2))+2\cos(\pi n( a -1))\Big) \nonumber \\
	&\hspace{10mm}+\Big(4(-1)^l N^{(+)}_{f}-2N^{s(+)}_f\Big)\cos(\pi n a ) \nonumber \\
	&\hspace{10mm}+\left.\Big(4(-1)^l N^{(-)}_{f}-2N^{s(-)}_f\Big)\cos(\pi n( a -1))\right]\,.
	\label{finiteTpt}
	\end{align}
The factor $(-1)^l$ comes from the anti-periodicity of fermions.
We can rewrite Eq.~\eqref{finiteTpt} as
	\begin{align}
	V^{T\ne0}_\mathrm{eff}( a ,T)&=2C\sum_{l=1}^{\infty}\sum_{n=1}^{\infty}\frac{1}{[n^2+l^2/(2\pi RT)^2]^{5/2}}\left[\Big(-3+4(-1)^l N^{(+)}_\mathrm{ad}\Big)\Big(\cos(2\pi n a )+2\cos(\pi n a )\Big)\right. \nonumber \\
	&\quad+4(-1)^l N^{(-)}_\mathrm{ad}\Big(\cos(2\pi n( a -1/2))+2\cos(\pi n( a -1))\Big) \nonumber \\
	&\quad+\Big(4(-1)^l N^{(+)}_{f}-2N^{s(+)}_f\Big)\cos(\pi n a ) \nonumber \\
	&\quad+\left.\Big(4(-1)^l N^{(-)}_{f}-2N^{s(-)}_f\Big)\cos(\pi n( a -1))\right]\,.
	\label{finiteTpt2}
	\end{align}
Finally, the total effective potential involving the temperature is given by
	\begin{align}
	V_\mathrm{eff}( a ,T)=V^{T=0}_\mathrm{eff}( a )+V^{T\ne0}_\mathrm{eff}( a ,T)\,.
	\label{totalVeff}
	\end{align}
In our numerical calculation, we use $V_{\text{eff}}(a,T)-V_{\text{eff}}(0,T)$ to take the potential value at $a=0$ as the origin.
Meanwhile, we can change the parameter $a$ in the potential to the scalar field value $\phi$ by $a=g_4R\phi$, as $V_\mathrm{eff}(a, T)\rightarrow V_\mathrm{eff}(\phi, T)$.

\section{Gravitational waves from first-order phase transitions}
\label{sec:3}
In this section, we will review GW spectrums from first-order phase transitions.
When the first-order phase transition occurs, bubbles appear in the universe and expand until colliding each other.
The nucleation temperature $T_{\rm nuc}$, where a bubble nucleates in one Hubble radius, is defined by $S_3/T\sim140$~\cite{McLerran:1990zh, Dine:1991ck, Anderson:1991zb}.
We can calculate the three-dimensional Euclidean action $S_3$ as
\begin{align}
S_3=\int d^3x\left[
\frac{1}{2}(\partial_\mu \phi)^2+V(\phi)
\right]\,,
\end{align}
with the scalar field $\phi$ and the potential $V$.
The shape of $\phi$ as a function of the bubble radius coordinate $r$ can be derived from the bounce solution. The solution has the $\mathcal{O}(3)$ symmetry at very high temperature~\cite{Linde:1981zj}, hence we can obtain the solution from the Euclidean equation of motion,
\begin{align}
\frac{d^2\phi}{dr^2} + \frac{2}{r}\frac{d\phi}{dr}=\frac{dV}{d\phi}\,,
\end{align}
where the boundary conditions are
\begin{align}
\underset{r \rightarrow \infty}{\mathrm{lim}} \phi(r) = 0\,, \quad
\left.\frac{d\phi}{dr}\right|_{r=0} = 0\,.
\end{align}

The GWs from the first-order phase transition arise from three sources.
Those are the collision of bubbles~\cite{Kosowsky:1991ua, Kosowsky:1992rz, Kosowsky:1992vn, Kamionkowski:1993fg, Caprini:2007xq,Huber:2008hg}, the sound waves in the plasma surrounding the bubble walls~\cite{Hindmarsh:2013xza, Hindmarsh:2015qta}, and magnetichydrodynamic (MHD) turbulence in the plasma~\cite{Caprini:2006jb, Kahniashvili:2008pf, Kahniashvili:2008pe, Kahniashvili:2009mf, Caprini:2009yp, Binetruy:2012ze}.
Each contribution to the GW spectrum can be written by parameters like $\alpha$ and $\tilde\beta$.
The phase transition strength $\alpha$ is
\begin{align}
\alpha \equiv& \frac{\epsilon(T_{\rm nuc})}{\rho_{\rm rad}(T_{\rm nuc})}\,,\label{eq:alpha}
\end{align}
with $\epsilon(T_{\rm nuc})\equiv(\Delta V-(1/4)T(\partial\Delta V/\partial T))|_{T=T_{\rm nuc}}$ and the radiation density $\rho_{\rm rad}(T)\equiv\pi^2g_*T^4/30$. 
\footnote{Conventionally, the phase transition strength was defined by using the difference of the enthalpy.
Currently, the standard way is to apply the pseudotrace to $\epsilon$ of the phase transition strength instead of the difference of the enthalpy. 
See Refs. \cite{Giese:2020rtr,Giese:2020znk}.
}
The index $\Delta V$ denotes the potential difference between the two minima for the first-order transition.
The effective relativistic degrees of freedom $g_*$ is 106.75 in the SM before the SSB. 
The $\tilde\beta$ is defined by $\beta/H(T_{\rm nuc})$ with the inverse duration of the phase transition $\beta$ and the Hubble rate at the nucleate temperature $H(T_{\rm nuc})$.
We can write the parameter $\tilde \beta$ as
\begin{align}
\tilde \beta& = \left.T \frac{d}{dT}\left(\frac{S_3}{T}\right)\right|_{T=T_{\rm nuc}}\, .\label{eq:beta}
\end{align}
From numerical simulations and analytic estimations, the GW spectrum for each source can be written as~\cite{Huber:2008hg, Hindmarsh:2015qta, Caprini:2009yp, Binetruy:2012ze}, 
\begin{align}
\Omega_{\varphi}h^2=&~
1.67\times 10^{-5}
\tilde\beta^{-2}
\left(\frac{\kappa_{\varphi} \alpha}{1+\alpha}\right)^2
\left(\frac{100}{g_{*}}\right)^{\frac{1}{3}}\left(\frac{0.11v^3_{w}}{0.42+v_w^2 }\right)\frac{3.8(f/f_{\varphi})^{2.8}}{1+2.8(f/f_{\varphi})^{3.8}}\,,
\label{eq:gw1}\\
\Omega_{\mathrm{sw}}h^2=&~
2.65\times 10^{-6}
\tilde\beta^{-1}
\left(\frac{\kappa_\text{sw} \alpha}{1+\alpha}\right)^2\left(\frac{100}{g_{*}}\right)^{\frac{1}{3}} v_{w}(f/f_{\mathrm{sw}})^3\left(\frac{7}{4+3(f/f_{\mathrm{sw}})^2}\right)^{\frac{7}{2}}\,,
\label{eq:gw2}\\
\Omega_{\mathrm{turb}}h^2=&~
3.35\times 10^{-4}
\tilde\beta^{-1}
\left(\frac{\kappa_\text{turb}\alpha}{1+\alpha}\right)^{\frac{3}{2}}\left(\frac{100}{g_{*}}\right)^{\frac{1}{3}}
v_{w}\frac{(f/f_{\mathrm{turb}})^3}{[
1+(f/f_{\mathrm{turb}})]^{\frac{11}{3}}(1+8\pi f/h_n)}\,,
\label{eq:gw3}
\end{align}
with bubble wall velocity $v_w$ and the frequency $f$.
The value of the inverse Hubble time at $T_{\rm nuc}$ redshifted today $h_n$ is given by
\begin{align}
h_n = 1.65 \times 10^{-5} \mathrm{Hz}~\left( \frac{T_{\rm nuc}}{100~\mathrm{GeV}}\right)\left( \frac{g_{*}}{100}\right)^{1/6}\,.
\end{align}
We denote the fraction of vacuum energy into each source as $\kappa_\varphi$, $\kappa_{\mathrm{sw}}$, and $\kappa_{\rm turb}$, which are estimated as~\cite{Kamionkowski:1993fg, Espinosa:2010hh, Hindmarsh:2015qta}
\begin{align}
\kappa_\varphi \approx&~ \frac{1}{1+0.715\alpha}\left(0.715\alpha+\frac{4}{27}\sqrt{\frac{3\alpha}{2}}\right)\,,\\
\kappa_{\mathrm{sw}} \approx&~ \frac{\alpha}{0.73+0.083\sqrt{\alpha}+\alpha}\,,\\
\kappa_{\rm turb} \approx&~ 0.1 \kappa_{\mathrm{sw}}\,.
\end{align}
The peak frequency for each GW contribution is respectively obtained as~\cite{Huber:2008hg, Hindmarsh:2015qta, Caprini:2009yp, Binetruy:2012ze}
\begin{align}
f_{\varphi} =&~ 1.65 \times 10^{-5} \mathrm{Hz}~
\tilde\beta
\left(\frac{0.62}{1.8-0.1 v_w+v^2_w }\right)\left( \frac{T_{\rm nuc}}{100~\mathrm{GeV}}\right)\left( \frac{g_{*}}{100}\right)^{\frac{1}{6}}\,,\\
 f_{\mathrm{sw}} =&~ 1.9 \times 10^{-5} \mathrm{Hz}~
v_w ^{-1}
\tilde\beta
 \left( \frac{T_{\rm nuc}}{100~\mathrm{GeV}}\right)\left( \frac{g_{*}}{100}\right)^{\frac{1}{6}}\,,\\
f_{\mathrm{turb}} =&~ 2.7 \times 10^{-5} \mathrm{Hz}~
v_w ^{-1}
\tilde\beta
 \left( \frac{T_{\rm nuc}}{100~\mathrm{GeV}}\right)\left( \frac{g_{*}}{100}\right)^{\frac{1}{6}}\,.
\end{align}
In the following, we calculate the GW spectrum as the combination of the three contributions,
\begin{align}
\Omega_\text{GW} h^2 =
\Omega_\varphi h^2 +
\Omega_\text{sw} h^2 + 
\Omega_\text{turb} h^2\,.
\label{GW-Omega}
\end{align}

\section{Numerical results} \label{sec:4}
In this section, we show relations between higher-dimensional gauge theories and GW spectrums yielded from them.
Following Refs.\cite{Haba:2004qf, Maru:2005jy}, we investigate two cases.

\subsection{Case 1}
Case 1 is simply chosen as
	\begin{align}
	N^{(+)}_f=3,\quad(\text{otherwise})=0\,.
	\end{align}
The effective potential at zero temperature is minimized at $a=a_0=1$, and the gauge symmetry is broken into $SU(2)\times U(1)\rightarrow U'(1)\times U(1)$.
\begin{figure}[t]
\begin{center}
\includegraphics[width=8cm]{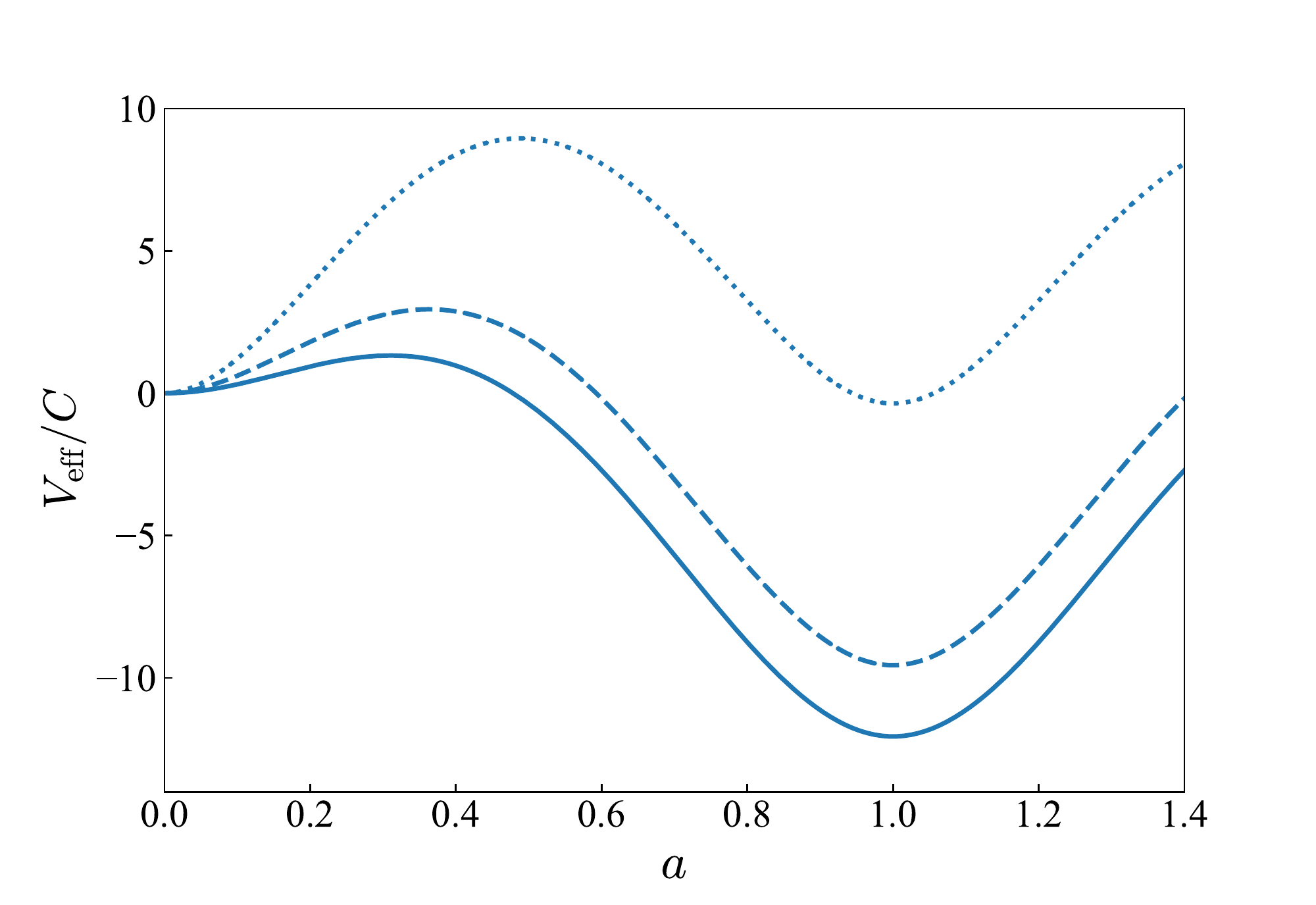}
\end{center}
\caption{
Shapes of the effective potential in case 1 for $RT=0$ (solid), 0.093 (dashed), and 0.15 (dotted).
}
\label{fig:Veff_case1}
\end{figure}
Fig.\ref{fig:Veff_case1} shows shapes of the effective potential in case 1 for $RT=0$ (solid), 0.093 (dashed), and 0.15 (dotted).
The local minima locate at $a=0$ and $1$ for all temperatures.
Beginning from large $RT$ like one to the low value, the potential goes bottom as the temperature decrease, and 
two minima at $a=0$ and 1 degenerate at $RT\simeq0.15$. We denote the temperature as the critical temperature $T_c$, which can be calculated as
\begin{align}
T_c\simeq \frac{1}{R}\times0.15\,.
\end{align}
After the degeneration, the minimum at $a=1$ becomes the global minimum for $RT<0.15$.
We can obtain the bubble nucleate temperature as, e.g. $RT_{\rm nuc}\simeq0.093$ for $(R, g_4)=(10^{-3}~{\rm GeV}^{-1}, 3)$ from the condition $S_3/T\sim140$ in Sec.~\ref{sec:3}.
In that case, $T_c\simeq150$ GeV and $T_{\rm nuc}\simeq93$ GeV.

\begin{figure}[t]
\begin{center}
\includegraphics[width=8cm]{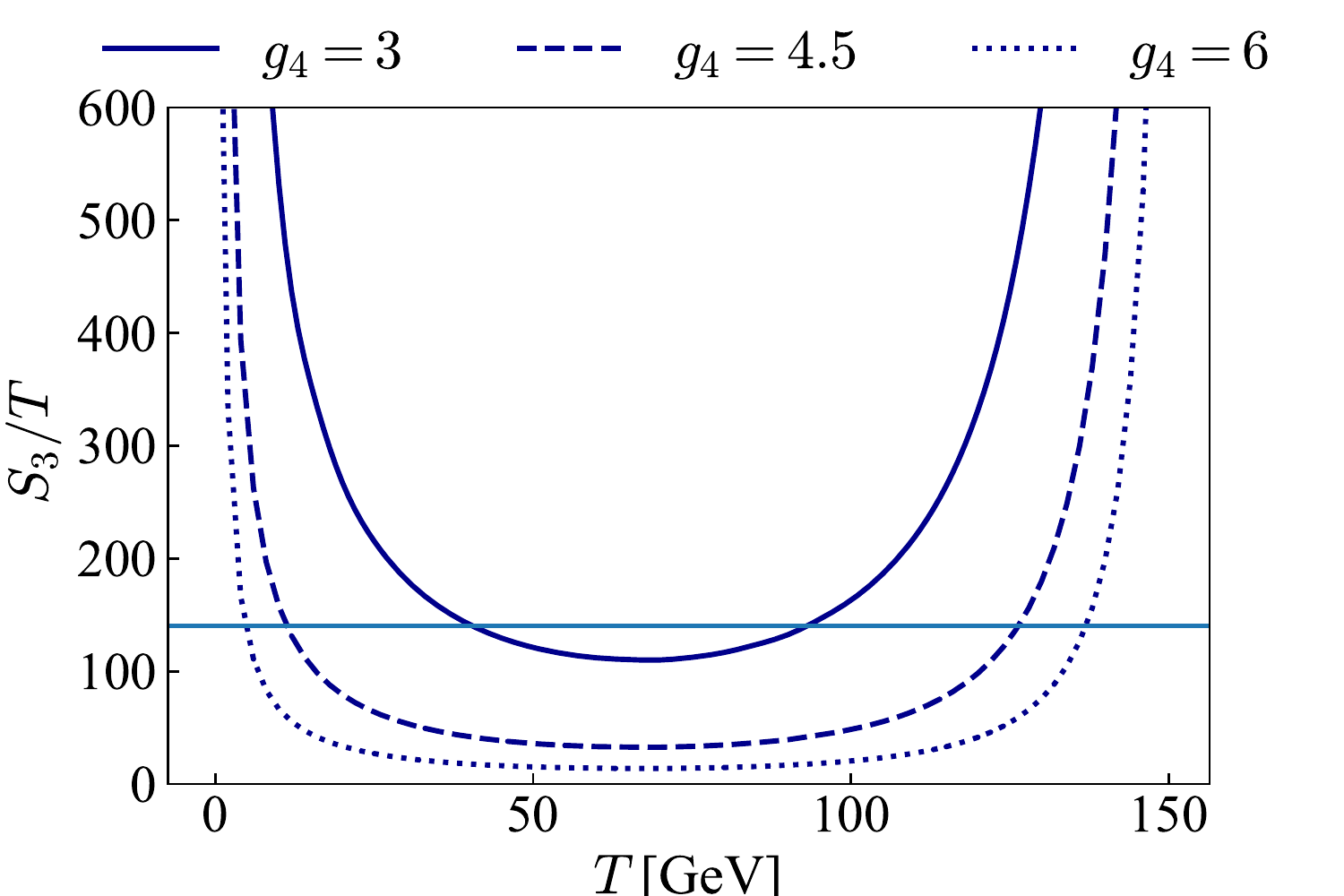}
\caption{
Action $S_3/T$ as a function of the temperature $T$ in case 1 with $R=10^{-3}$ GeV$^{-1}$. The solid, dashed, and dotted lines correspond to $g_4=3$, 4.5, and 6, respectively.
}
\label{fig:S3overT_case1}
\end{center} 
\end{figure}

To see the dependence of $g_4$ on the phase transition, we show the action $S_3/T$ as a function of the temperature for different values of $g_4$ in Fig.~\ref{fig:S3overT_case1}. To demonstrate specifically, we start from $R=10^{-3}$ GeV$^{-1}$ as a example.
We calculate the action using CosmoTransitions~\cite{Wainwright:2011kj} and confirmed the result by FindBounce~\cite{Guada:2020xnz}.
The solid, dashed, and dotted lines in Fig.~\ref{fig:S3overT_case1} represent results with $g_4=3$, 4.5, and 6, respectively. 
We take $g_4\gtrsim 2.8$ because we numerically find that $S_3/T$ does not reach under 140 for $g_4\lesssim2.8$.
If $S_3/T$ is always larger than 140, the SSB would not occur, and the vacuum would be trapped at the origin even at zero temperature.
In Fig.~\ref{fig:S3overT_case1}, the action $S_3/T$ for $g_4=3$ reaches 140 at $T_{\rm nuc}\simeq92.9$ GeV when starting from high temperature.
For $T<70$ GeV, $S_3/T$ starts to increase until zero temperature. It is because the potential barrier between two minima remains at zero temperature.
We can also see the nucleate temperature $T_{\rm nuc}$ increases and approaches $T_c$ as $g_4$ becomes larger like $T_{\rm nuc}\simeq92.9, 126,$ and 137 GeV for $g_4=3,4.5$, and 6, respectively.
Note that since we fix $R$ as $10^{-3}$, $T_c$ is also constant as $T_c\simeq150$ GeV.
For the larger $g_4$, the phase transition becomes easy to occur because it makes the field distance $v$ between the two minima smaller, while the height of the potential barrier and the depth of the global minimum $\Delta V$ remain. We can see the relation between $g_4$ and $v$ from Eq.~\eqref{eq:relation}.

\begin{figure}[t]
 \begin{center}
\includegraphics[width=7.5cm]{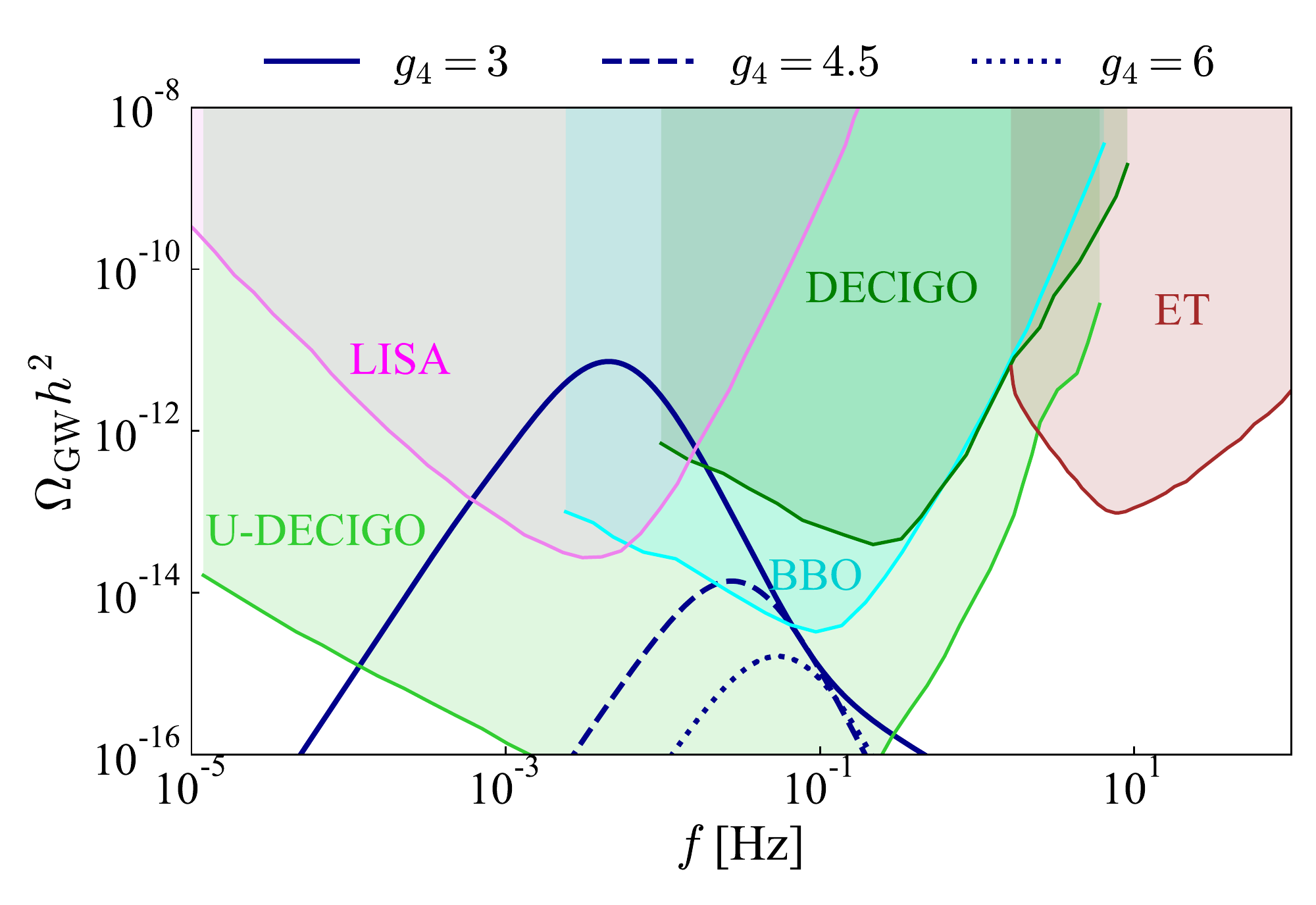}
\includegraphics[width=7.5cm]{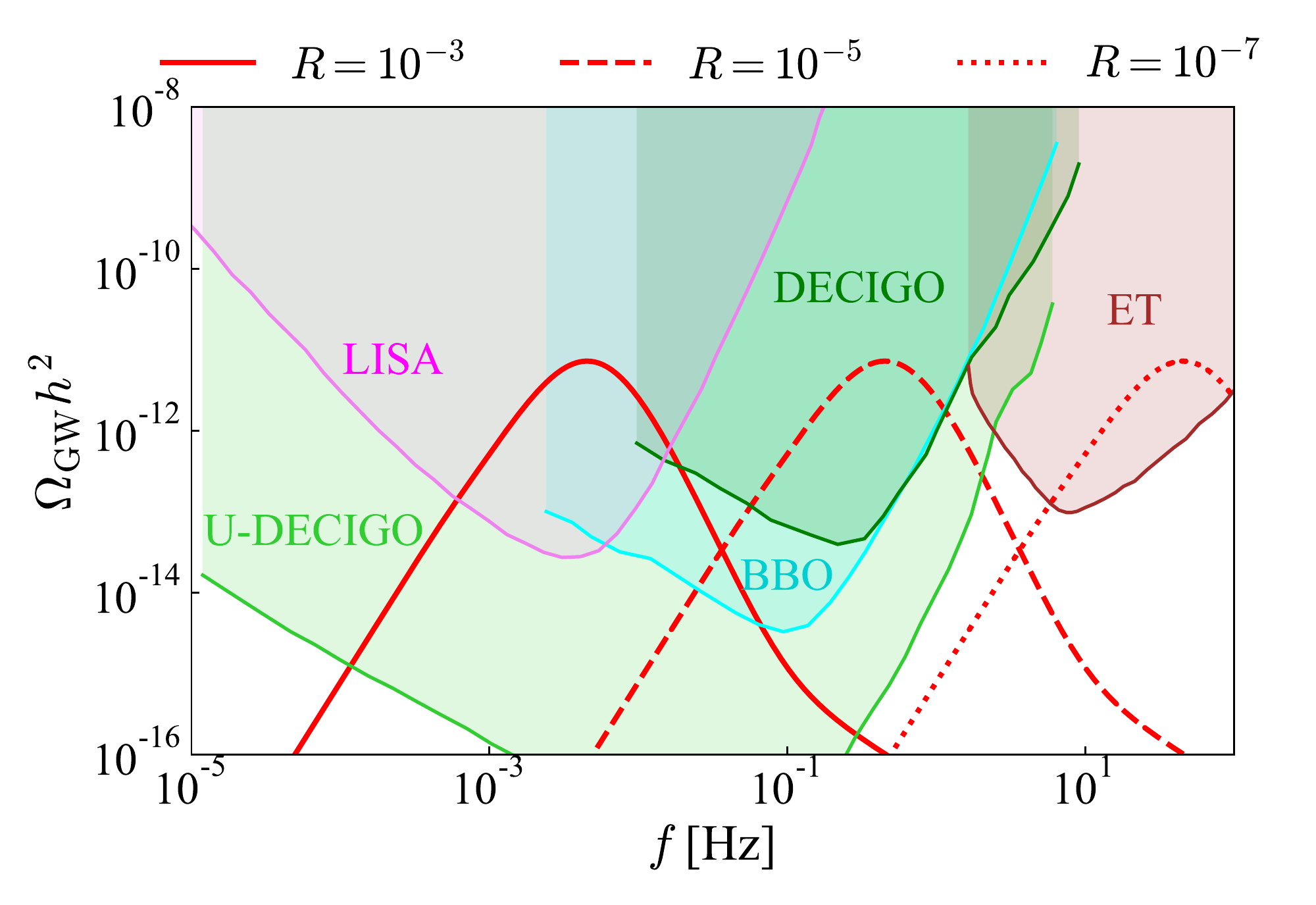}
\caption{
GW spectrums in case 1 for $g_4=3, 4.5$, 6 with $R=10^{-3}$ GeV$^{-1}$ (left), and $R=10^{-3}, 10^{-5}$, and $10^{-7}$ GeV$^{-1}$ with $g_4=3$ (right).
The colored regions indicate observed areas by future interferometers.
}
\label{fig:GW_case1}
 \end{center}
\end{figure}

From now on, we show the GW spectrum from the first-order phase transition occurring in the higher-dimensional gauge theory. 
The left panel of Fig.~\ref{fig:GW_case1} represents the GW spectrums for the same values of $g_4$ and $R$ as in Fig~\ref{fig:S3overT_case1}. 
The colored regions indicate observable areas in future space-based interferometers, which are LISA~\cite{Caprini:2015zlo, LISA:2017pwj, Caprini:2019egz}, DECIGO~\cite{Seto:2001qf, Kawamura:2011zz, Kawamura:2020pcg}, Big Bang Observer (BBO)~\cite{Corbin:2005ny}, and Ultimate DECIGO (U-DECIGO)~\cite{Kudoh:2005as}, and the ground-based interferometer, ET~\cite{Punturo:2010zz}.
We can see that the peak of the spectrum moves to the bottom right as $g_4$ increases.
Moving the high frequency implies that $T_{\rm nuc}$ becomes higher, and the temperature derivative of $S_3/T$ becomes larger. We can see the features in Fig.~\ref{fig:S3overT_case1}.
The reduction of the GW energy density comes from the decrease of the strength of the phase transition $\alpha$.
It is because $T_{\rm nuc}$ approaches $T_c$ as $g_4$ increases as shown in Fig.~\ref{fig:S3overT_case1}.
In this case, the higher $T_{\rm nuc}$ reduces the difference of the potential $\Delta V$ when the phase transition occurs.
The small $\Delta V$ leads the small latent heat $\epsilon$, and the high $T_{\rm nuc}$ indicates high radiation density $\rho_{\rm rad}$.
They lead to small $\alpha$.
Hence, the energy density of the GW spectrum decreases as $g_4$ rises.
Since we numerically find that GW spectrum for $g_4\simeq4$ cannot be detected by LISA and $S_3/T$ does not reach under 140 for $g_4\lesssim2.8$, LISA can observe the region $2.8\lesssim g_4\lesssim4$ with $R=10^{-3}$ GeV$^{-1}$, while BBO and U-DECIGO can investigate $g_4\gtrsim2.8$ for the SSB.

Until now, we consider the phase transitions with the compact scale $R=10^{-3}$ GeV$^{-1}$. 
However, scales of higher-dimensional theories $1/R$ can be higher. In the right panel of Fig.~\ref{fig:GW_case1}, we show the GW spectrums when changing $R$ with the fixed coupling $g_4=3$.
We can see that the peaks of spectrums slide right side as $R$ decreases while keeping the spectrum shapes. It is because the GW parameters, $\alpha$ and $\tilde{\beta}$, do not depend on $R$.
The values for $\alpha$ and $\tilde{\beta}$ are stable since the potential shape changes by similarity for changing $R$.
For example, the latent heat $\epsilon$ and the radiation density $\rho_{\rm rad}$ alter at the same rate for changing $R$, hence the ratio $\alpha$ between them remains the same value.
As a result of the stability of $\alpha$ and $\tilde{\beta}$, the changes of GW spectrums depend only on $T_{\rm nuc}$. The nucleate temperature $T_{\rm nuc}$ increases with the scale $1/R$, and the higher $T_{\rm nuc}$ make the GW peak frequency larger as in the right panel of Fig.~\ref{fig:GW_case1}. 
Observable regions of $R$ and the VEV $v$ of the extra dimensional gauge component for each interferometer are shown in Tab.~\ref{tab:1}.
The compact scale $R$ is inversely proportional to the scalar VEV $v$ as in Eq.~\eqref{eq:relation}. 
Through the combination of the interferometers, the scale of the higher-dimensional gauge theory can be investigated up to $\mathcal{O}(10^{7})$ GeV. 

\begin{table}[t]
\caption{
Observable scales of the compact scale $R$ and the VEV $v$ by the future interferometers.
}
\centering
\begin{tabular}{c|cc}
 & $R$ [GeV$^{-1}$] & $v$ [GeV] \\ \hline
LISA & $10^{-4}$~--~$10^{-1}$ & $10^{1}$~--~$10^{3}$ \\
DECIGO & $10^{-6}$~--~$10^{-3}$ & $10^{3}$~--~$10^{5}$ \\
ET & $10^{-7}$~--~$10^{-6}$ & $10^{5}$~--~$10^{7}$ \\
\end{tabular}
\label{tab:1}
\end{table}

\subsection{Case 2}
Case 2 is chosen as
	\begin{align}
	N^{(+)}_\mathrm{ad}=2,~~~N^{(-)}_\mathrm{ad}=0,~~~N^{(+)}_{f}=0,~~~N^{(-)}_{f}=8,~~~N^{s(+)}_{f}=4,~~~N^{s(-)}_{f}=2\,,
	\label{case2}
	\end{align}
following Refs.\cite{Haba:2004qf, Maru:2005jy}.
The effective potential at zero temperature is minimized at $a_0\simeq0.058$, and the gauge symmetry is broken into $SU(2)\times U(1)\rightarrow U(1)$.
\begin{figure}[t]
\begin{center}
\includegraphics[width=8cm]{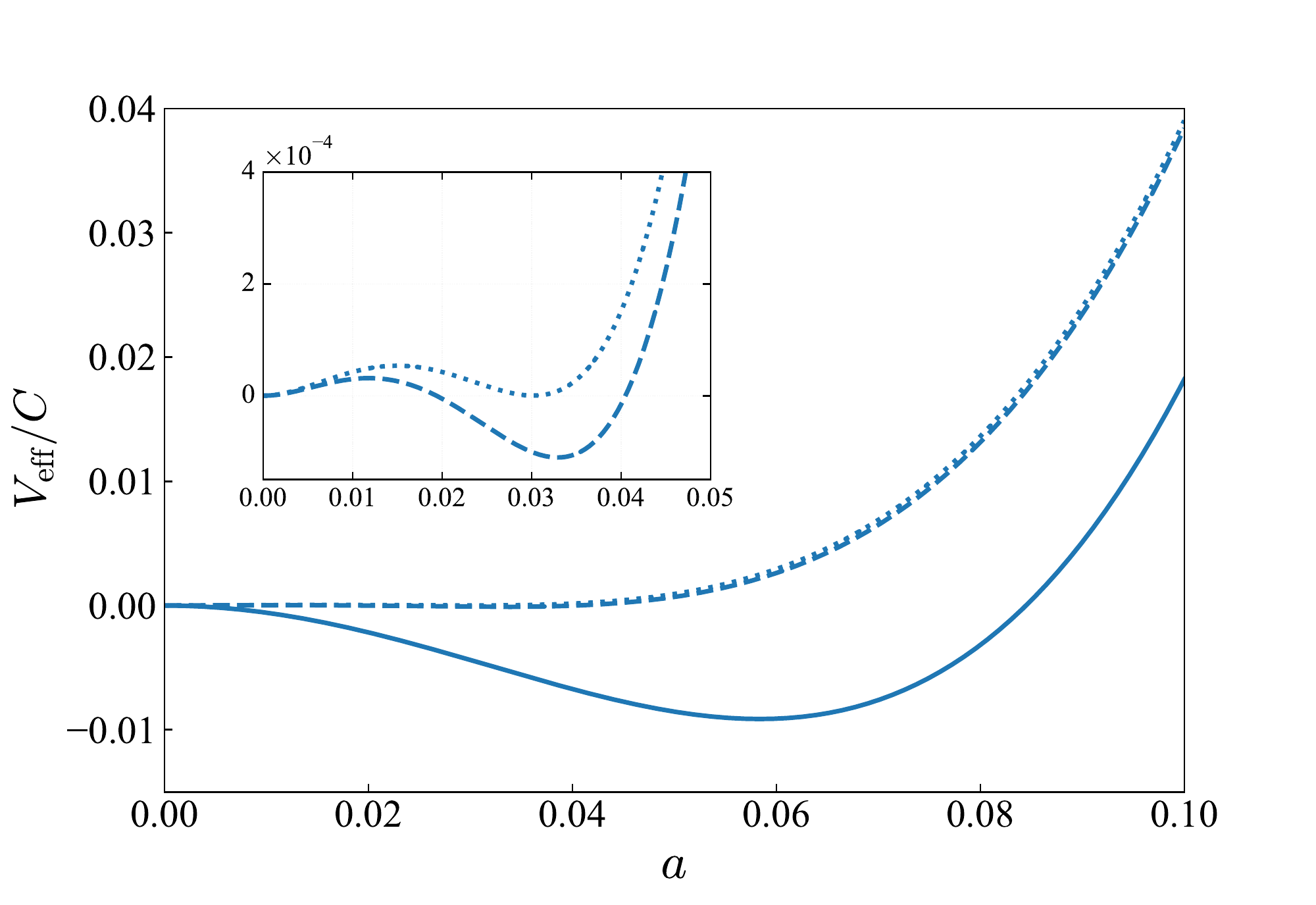}
\end{center}
\caption{
Shapes of the effective potential in case 2 for $RT=0$ (solid), 0.0246 (dashed), and 0.0254 (dotted).}
\label{fig:Veff_case2}
\end{figure}
Fig.\ref{fig:Veff_case2} shows shapes of the effective potential in case 2 for $RT=0$ (solid), 0.0246 (dashed), and 0.0254 (dotted).
The critical temperature is numerically given by
\begin{align}
T_c\simeq \frac{1}{R}\times0.0254\,,
\end{align}
from the dotted line in Fig.\ref{fig:Veff_case2}.
We can see the potential barrier at $T_c$ is tiny compared with that for case 1 in Fig.~\ref{fig:Veff_case1}. It would be because the scale of $a$ at the minimum for $a\neq0$ is much smaller than that in case 1. The potential barrier disappears at zero temperature, while it remains in case 1.

\begin{figure}[t]
\begin{center}
\includegraphics[width=8cm]{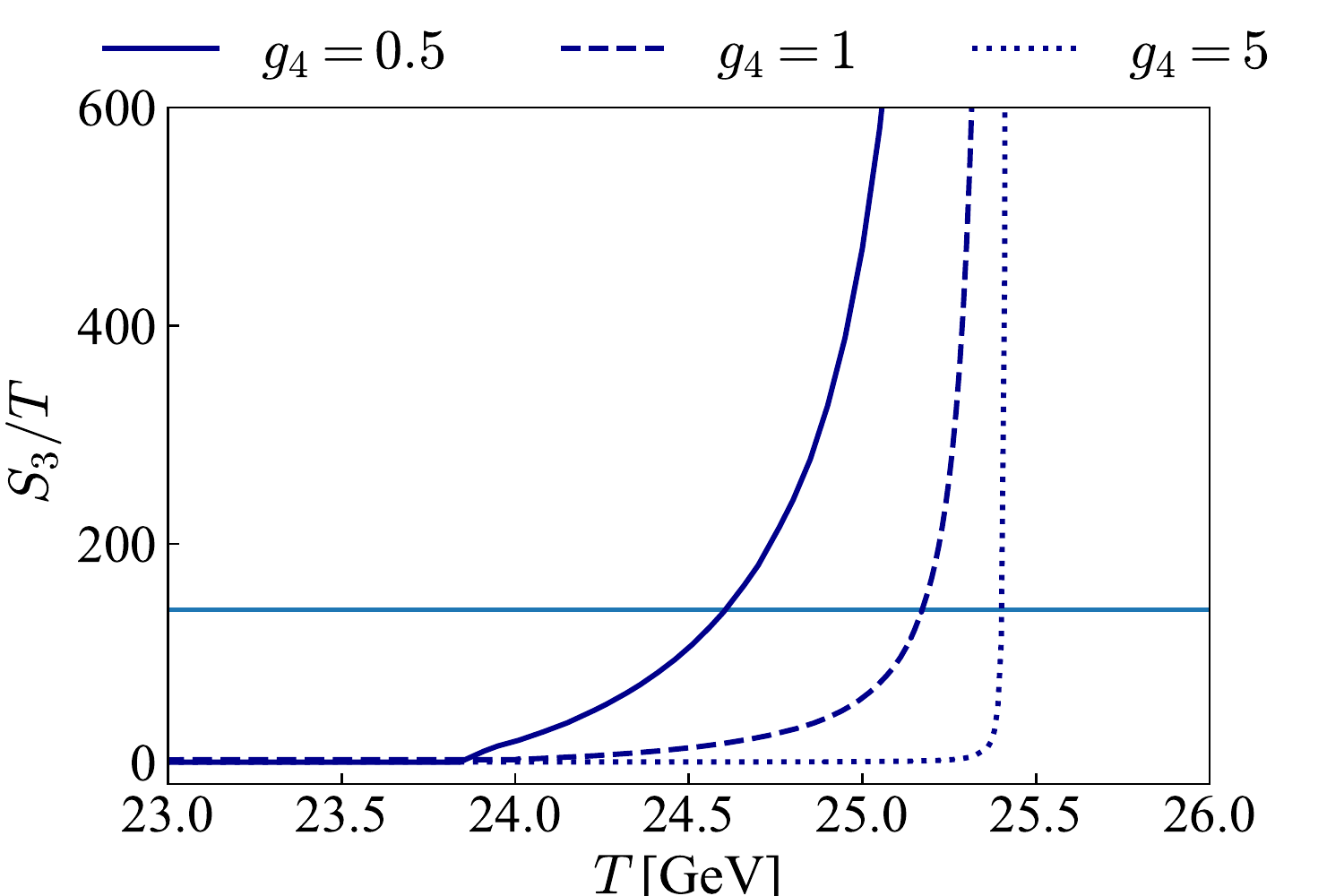}
\caption{
Action $S_3/T$ as a function of the temperature $T$ in case 2 with $R=10^{-3}$ GeV$^{-1}$.
The dashed, solid, and dotted lines correspond to $g_4=0.5$, 1, and 5, respectively.
}
\label{fig:S3overT_case2}
\end{center} 
\end{figure}
\begin{figure}[t]
 \begin{center}
\includegraphics[width=7.5cm]{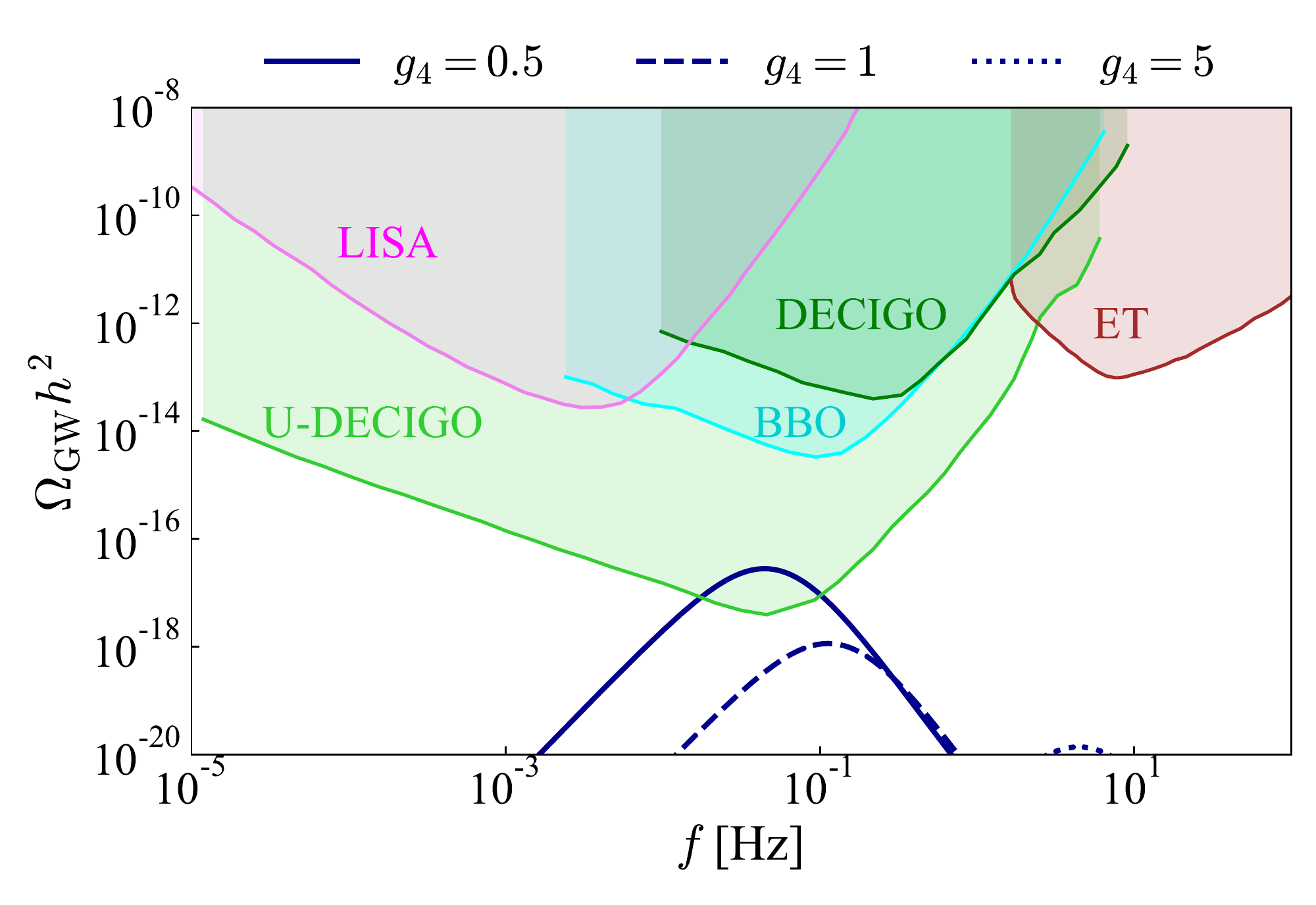}
\includegraphics[width=7.5cm]{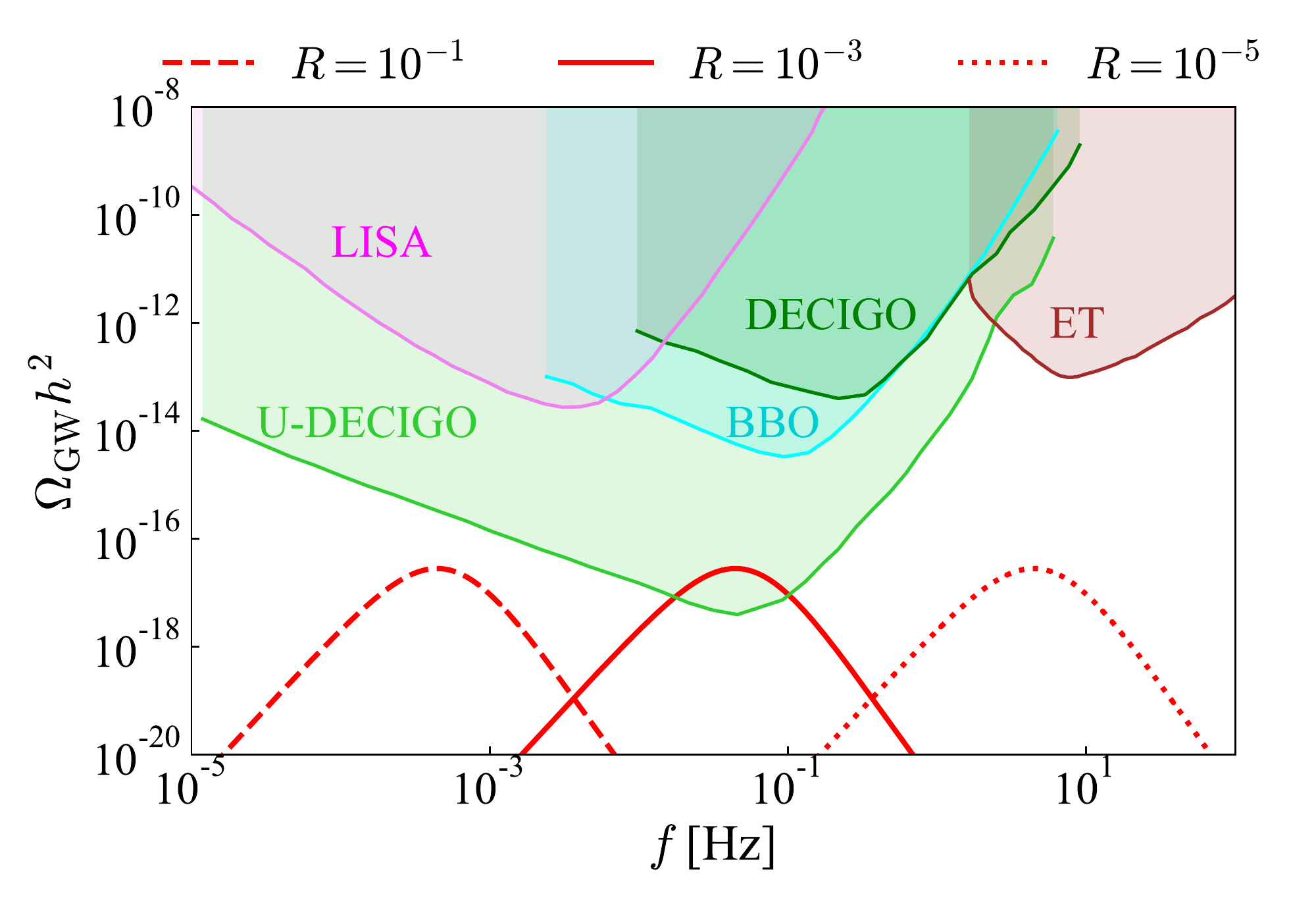}
\caption{
GW spectrums in case 2 for $g_4=0.5, 1$, and 5 with $R=10^{-3}$ GeV$^{-1}$ (left), and $R=10^{-1}, 10^{-3}$, and $10^{-5}$ GeV$^{-1}$ with $g_4=0.5$ (right).
}
\label{fig:GW_case2}
 \end{center}
\end{figure}

To see $g_4$ dependence on the phase transition as in case 1, we show the action $S_3/T$ as the function of $T$ for case 2 in Fig.~\ref{fig:S3overT_case2}. As the same as case 1, we take $R=10^{-3}$ GeV$^{-1}$ for an example. The solid, dashed, and dotted line indicates results for $g_4=0.5, 1,$ and 5.
We can see the same feature, which is that the nucleate temperature $T_{\rm nuc}$ becomes larger and approaches $T_c$ as $g_4$ increases. The values of $T_{\rm nuc}$ are 24.6 GeV, 25.2 GeV, and 25.4 GeV for $g_4=0.5, 1,$ and 5, respectively.
The shift of $T_{\rm nuc}$ for changing $g_4$ is slight because the potential barrier is much smaller than the scale of $V_{\rm eff}$ as shown in Fig.~\ref{fig:Veff_case2} and the barrier change significantly by the slight shift of $T$. Besides, contrary to case 1, $S_3/T$ reaches zero and remains until zero temperature because there is no potential barrier anymore. Since the potential vanishes, $S_3/T$ always reaches 140 for an arbitrary value of $g_4$.

The left panel of Fig.~\ref{fig:GW_case2} represents the GW spectrums for the same values of $g_4$ as in Fig.~\ref{fig:S3overT_case2}.
We can see the same feature for changing $g_4$ as in the left panel of Fig.~\ref{fig:GW_case1}, while the GW energy density is smaller. Since the potential barrier is tiny, the potential difference $\Delta V$ at $T_{\rm nuc}$ becomes small. It leads to the small $\alpha$ and GW energy density.
The interferometer U-DECIGO is possible to investigate the region $g_4<1$. 
In the right panel of Fig.~\ref{fig:GW_case2} shows the GW spectrums for $R=10^{-1}, 10^{-3}$, and $10^{-5}$ GeV$^{-1}$ with $g_4=0.5$.
As the same as in case 1, the spectrum slides for changing $R$ while keeping the shape. The observable regions by U-DECIGO are $10^{-4}~\mathrm{GeV}^{-1} < R\lesssim10^{-2}~\mathrm{GeV}^{-1}$ and $10^{1}~\mathrm{GeV} \lesssim v < 10^{2}~\mathrm{GeV}$. 

\section{Conclusion} \label{sec:5}
In this paper, we have considered the five-dimensional $SU(3)$ gauge theories with the orbifold $S^1/Z_2$ and discuss its verifiability in future GW observations.
At first, we formulated the thermal effective potential, which is possible to cause the SSB. 
We perform numerical calculations in the two cases of field choices where the first-order phase transitions can occur. 
In case 1, we introduced fermions in fundamental representation.
The symmetry breaking pattern is $SU(2)\times U(1)\rightarrow U'(1)\times U(1)$. 
The characteristic point is that the potential barrier remains at zero temperature.
In case 2, where the potential barrier vanishes at a certain temperature, we introduced matter fields as Eq.\eqref{case2}.
The symmetry breaking pattern is $SU(2)\times U(1)\rightarrow U(1)$.
This case is similar to the electroweak symmetry breaking pattern.

Through our numerical calculation,
we revealed the relations between the GW spectrums and the distinctive parameters, which are the four-dimensional gauge coupling $g_4$ and the compact scale $R$ for the orbifold.
The main results in case 1 are Fig.~\ref{fig:GW_case1}.
Since $S_3/T$ does not reach under 140 for $g_4\lesssim2.8$ in Fig.~\ref{fig:S3overT_case1},
we found that LISA can observe the region $2.8\lesssim g_4\lesssim4$ with $R=10^{-3}$ GeV$^{-1}$.
The main results in case 2 are also Fig.~\ref{fig:GW_case2}.
In this case, the interferometer U-DECIGO is possible to investigate the region $g_4<1$.
In addition, we found that the observable regions by U-DECIGO are $10^{-4}~\mathrm{GeV}^{-1} < R\lesssim10^{-2}~\mathrm{GeV}^{-1}$.

In both cases, larger $g_4$ makes the GW peak frequency higher and the GW energy density smaller. 
The phenomenon can be explained by the change of the behavior of $S_3/T$ for changing $g_4$. 
On the other hand, $R$ is irrelevant to the shape of the GW spectrum, while it controls the GW peak frequency.
These features are general properties of GW spectrum from phase transitions. 
To be more specific in higher dimensional gauge theories, the independence of $R$ on the shape of the GW spectrums comes from the similarity of the potential change for shifting $R$. 
Regarding the difference between the two cases, the important factors are the scale of the potential barrier between the two minima and the disappearance of the barrier.
If the potential barrier at the phase transition is large enough, the produced GW energy density would be huge. In that case, the verifiability in the GW observations would be hopeful, and we can investigate wide ranges of the parameters by the combination of the observations.
Meanwhile, if the potential barrier remains even at zero temperature, the SSB would occur only for large enough $g_4$.
The coupling $g_4$ also controls the GW energy density, hence the value of $g_4$ would be essential for discussing the verifiability of the GW observations.

As an application of our results, we can consider the gauge-Higgs unification (GHU), where the zero-mode of the scalar field induced from extra components of a higher-dimensional gauge field is identified with the Higgs boson.
To apply our results to gauge-Higgs unification, we must identify the zero-modes of extra components of higher-dimensional gauge fields as Higgs fields and realize the masses of W gauge boson and Higgs boson.
In order to realize these masses, the GHU with matter fields in higher representations, e.g., $\textbf{10}$ and $\textbf{15}$ representations, has been studied (see Refs.\cite{Scrucca:2003ra, Adachi:2018mby}).
In the future, we will come back to the relation between the GHU (or higher-dimensional gauge theories with the matter fields in higher representation) and the GW spectrum from the first-order phase transitions.

\section*{Acknowledgments}
We would like to thank Nobuhito Maru for his valuable discussions.
This work was supported by JSPS KAKENHI Grant No. JP22J15562 (TH), and JST SPRING, Grand No.~JPMJSP2135 (HS).

\let\doi\relax
\bibliographystyle{junsrt}
\bibliography{Ref}

\begin{thebibliography}{10}

\bibitem{Manton:1979kb}
N.~S. Manton.
\newblock {A New Six-Dimensional Approach to the Weinberg-Salam Model}.
\newblock {\em Nucl. Phys. B}, Vol. 158, pp. 141--153, 1979.

\bibitem{Arkani-Hamed:1998jmv}
Nima Arkani-Hamed, Savas Dimopoulos, and G.~R. Dvali.
\newblock {The Hierarchy problem and new dimensions at a millimeter}.
\newblock {\em Phys. Lett. B}, Vol. 429, pp. 263--272, 1998.

\bibitem{Randall:1999ee}
Lisa Randall and Raman Sundrum.
\newblock {A Large mass hierarchy from a small extra dimension}.
\newblock {\em Phys. Rev. Lett.}, Vol.~83, pp. 3370--3373, 1999.

\bibitem{Appelquist:2000nn}
Thomas Appelquist, Hsin-Chia Cheng, and Bogdan~A. Dobrescu.
\newblock {Bounds on universal extra dimensions}.
\newblock {\em Phys. Rev. D}, Vol.~64, p. 035002, 2001.

\bibitem{Hosotani:1983xw}
Yutaka Hosotani.
\newblock {Dynamical Mass Generation by Compact Extra Dimensions}.
\newblock {\em Phys. Lett. B}, Vol. 126, pp. 309--313, 1983.

\bibitem{Hosotani:1988bm}
Yutaka Hosotani.
\newblock {Dynamics of Nonintegrable Phases and Gauge Symmetry Breaking}.
\newblock {\em Annals Phys.}, Vol. 190, p. 233, 1989.

\bibitem{Hatanaka:1998yp}
Hisaki Hatanaka, Takeo Inami, and C.~S. Lim.
\newblock {The Gauge hierarchy problem and higher dimensional gauge theories}.
\newblock {\em Mod. Phys. Lett. A}, Vol.~13, pp. 2601--2612, 1998.

\bibitem{Kubo:2001zc}
Masahiro Kubo, C.~S. Lim, and Hiroyuki Yamashita.
\newblock {The Hosotani mechanism in bulk gauge theories with an orbifold extra
  space S**1 / Z(2)}.
\newblock {\em Mod. Phys. Lett. A}, Vol.~17, pp. 2249--2264, 2002.

\bibitem{Scrucca:2003ra}
Claudio~A. Scrucca, Marco Serone, and Luca Silvestrini.
\newblock {Electroweak symmetry breaking and fermion masses from extra
  dimensions}.
\newblock {\em Nucl. Phys. B}, Vol. 669, pp. 128--158, 2003.

\bibitem{Haba:2004qf}
Naoyuki Haba, Yutaka Hosotani, Yoshiharu Kawamura, and Toshifumi Yamashita.
\newblock {Dynamical symmetry breaking in gauge Higgs unification on orbifold}.
\newblock {\em Phys. Rev. D}, Vol.~70, p. 015010, 2004.

\bibitem{Maru:2006wa}
Nobuhito Maru and Toshifumi Yamashita.
\newblock {Two-loop Calculation of Higgs Mass in Gauge-Higgs Unification: 5D
  Massless QED Compactified on S**1}.
\newblock {\em Nucl. Phys. B}, Vol. 754, pp. 127--145, 2006.

\bibitem{Hosotani:2007kn}
Y.~Hosotani, N.~Maru, K.~Takenaga, and Toshifumi Yamashita.
\newblock {Two Loop finiteness of Higgs mass and potential in the gauge-Higgs
  unification}.
\newblock {\em Prog. Theor. Phys.}, Vol. 118, pp. 1053--1068, 2007.

\bibitem{Adachi:2018mby}
Yuki Adachi and Nobuhito Maru.
\newblock {Revisiting electroweak symmetry breaking and the Higgs boson mass in
  gauge-Higgs unification}.
\newblock {\em Phys. Rev. D}, Vol.~98, No.~1, p. 015022, 2018.

\bibitem{Panico:2005ft}
Giuliano Panico and Marco Serone.
\newblock {The Electroweak phase transition on orbifolds with gauge-Higgs
  unification}.
\newblock {\em JHEP}, Vol.~05, p. 024, 2005.

\bibitem{Maru:2005jy}
Nobuhito Maru and Kazunori Takenaga.
\newblock {Aspects of phase transition in gauge-Higgs unification at finite
  temperature}.
\newblock {\em Phys. Rev. D}, Vol.~72, p. 046003, 2005.

\bibitem{Maru:2006wx}
Nobuhito Maru and Kazunori Takenaga.
\newblock {Bulk Mass Effects in Gauge-Higgs Unification at Finite Temperature}.
\newblock {\em Phys. Rev. D}, Vol.~74, p. 015017, 2006.

\bibitem{Adachi:2019apm}
Yuki Adachi, Nobuhito Maru, and Nobuhito Maru.
\newblock {Strong First Order Electroweak Phase Transition in Gauge-Higgs
  Unification at Finite Temperature}.
\newblock {\em Phys. Rev. D}, Vol. 101, No.~3, p. 036013, 2020.

\bibitem{Funatsu:2021gnh}
Shuichiro Funatsu, Hisaki Hatanaka, Yutaka Hosotani, Yuta Orikasa, and Naoki
  Yamatsu.
\newblock {Electroweak and left-right phase transitions in
  SO(5)\texttimes{}U(1)\texttimes{}SU(3) gauge-Higgs unification}.
\newblock {\em Phys. Rev. D}, Vol. 104, No.~11, p. 115018, 2021.

\bibitem{Caprini:2015zlo}
Chiara Caprini, et~al.
\newblock {Science with the space-based interferometer eLISA. II: Gravitational
  waves from cosmological phase transitions}.
\newblock {\em JCAP}, Vol.~04, p. 001, 2016.

\bibitem{LISA:2017pwj}
Pau Amaro-Seoane, et~al.
\newblock {Laser Interferometer Space Antenna}.
\newblock 2 2017.

\bibitem{Caprini:2019egz}
Chiara Caprini, et~al.
\newblock {Detecting gravitational waves from cosmological phase transitions
  with LISA: an update}.
\newblock {\em JCAP}, Vol.~03, p. 024, 2020.

\bibitem{Seto:2001qf}
Naoki Seto, Seiji Kawamura, and Takashi Nakamura.
\newblock {Possibility of direct measurement of the acceleration of the
  universe using 0.1-Hz band laser interferometer gravitational wave antenna in
  space}.
\newblock {\em Phys. Rev. Lett.}, Vol.~87, p. 221103, 2001.

\bibitem{Kawamura:2011zz}
Seiji Kawamura, et~al.
\newblock {The Japanese space gravitational wave antenna: DECIGO}.
\newblock {\em Class. Quant. Grav.}, Vol.~28, p. 094011, 2011.

\bibitem{Kawamura:2020pcg}
Seiji Kawamura, et~al.
\newblock {Current status of space gravitational wave antenna DECIGO and
  B-DECIGO}.
\newblock {\em PTEP}, Vol. 2021, No.~5, p. 05A105, 2021.

\bibitem{Punturo:2010zz}
M.~Punturo, et~al.
\newblock {The Einstein Telescope: A third-generation gravitational wave
  observatory}.
\newblock {\em Class. Quant. Grav.}, Vol.~27, p. 194002, 2010.

\bibitem{Antoniadis:2001cv}
Ignatios Antoniadis, K.~Benakli, and M.~Quiros.
\newblock {Finite Higgs mass without supersymmetry}.
\newblock {\em New J. Phys.}, Vol.~3, p.~20, 2001.

\bibitem{McLerran:1990zh}
Larry~D. McLerran, Mikhail~E. Shaposhnikov, Neil Turok, and Mikhail~B.
  Voloshin.
\newblock {Why the baryon asymmetry of the universe is approximately 10**-10}.
\newblock {\em Phys. Lett. B}, Vol. 256, pp. 451--456, 1991.

\bibitem{Dine:1991ck}
Michael Dine, Patrick Huet, and Robert~L. Singleton, Jr.
\newblock {Baryogenesis at the electroweak scale}.
\newblock {\em Nucl. Phys. B}, Vol. 375, pp. 625--648, 1992.

\bibitem{Anderson:1991zb}
Greg~W. Anderson and Lawrence~J. Hall.
\newblock {The Electroweak phase transition and baryogenesis}.
\newblock {\em Phys. Rev. D}, Vol.~45, pp. 2685--2698, 1992.

\bibitem{Linde:1981zj}
Andrei~D. Linde.
\newblock {Decay of the False Vacuum at Finite Temperature}.
\newblock {\em Nucl. Phys. B}, Vol. 216, p. 421, 1983.
\newblock [Erratum: Nucl.Phys.B 223, 544 (1983)].

\bibitem{Kosowsky:1991ua}
Arthur Kosowsky, Michael~S. Turner, and Richard Watkins.
\newblock {Gravitational radiation from colliding vacuum bubbles}.
\newblock {\em Phys. Rev. D}, Vol.~45, pp. 4514--4535, 1992.

\bibitem{Kosowsky:1992rz}
Arthur Kosowsky, Michael~S. Turner, and Richard Watkins.
\newblock {Gravitational waves from first order cosmological phase
  transitions}.
\newblock {\em Phys. Rev. Lett.}, Vol.~69, pp. 2026--2029, 1992.

\bibitem{Kosowsky:1992vn}
Arthur Kosowsky and Michael~S. Turner.
\newblock {Gravitational radiation from colliding vacuum bubbles: envelope
  approximation to many bubble collisions}.
\newblock {\em Phys. Rev. D}, Vol.~47, pp. 4372--4391, 1993.

\bibitem{Kamionkowski:1993fg}
Marc Kamionkowski, Arthur Kosowsky, and Michael~S. Turner.
\newblock {Gravitational radiation from first order phase transitions}.
\newblock {\em Phys. Rev. D}, Vol.~49, pp. 2837--2851, 1994.

\bibitem{Caprini:2007xq}
Chiara Caprini, Ruth Durrer, and Geraldine Servant.
\newblock {Gravitational wave generation from bubble collisions in first-order
  phase transitions: An analytic approach}.
\newblock {\em Phys. Rev. D}, Vol.~77, p. 124015, 2008.

\bibitem{Huber:2008hg}
Stephan~J. Huber and Thomas Konstandin.
\newblock {Gravitational Wave Production by Collisions: More Bubbles}.
\newblock {\em JCAP}, Vol.~09, p. 022, 2008.

\bibitem{Hindmarsh:2013xza}
Mark Hindmarsh, Stephan~J. Huber, Kari Rummukainen, and David~J. Weir.
\newblock {Gravitational waves from the sound of a first order phase
  transition}.
\newblock {\em Phys. Rev. Lett.}, Vol. 112, p. 041301, 2014.

\bibitem{Hindmarsh:2015qta}
Mark Hindmarsh, Stephan~J. Huber, Kari Rummukainen, and David~J. Weir.
\newblock {Numerical simulations of acoustically generated gravitational waves
  at a first order phase transition}.
\newblock {\em Phys. Rev. D}, Vol.~92, No.~12, p. 123009, 2015.

\bibitem{Caprini:2006jb}
Chiara Caprini and Ruth Durrer.
\newblock {Gravitational waves from stochastic relativistic sources: Primordial
  turbulence and magnetic fields}.
\newblock {\em Phys. Rev. D}, Vol.~74, p. 063521, 2006.

\bibitem{Kahniashvili:2008pf}
Tina Kahniashvili, Arthur Kosowsky, Grigol Gogoberidze, and Yurii Maravin.
\newblock {Detectability of Gravitational Waves from Phase Transitions}.
\newblock {\em Phys. Rev. D}, Vol.~78, p. 043003, 2008.

\bibitem{Kahniashvili:2008pe}
Tina Kahniashvili, Leonardo Campanelli, Grigol Gogoberidze, Yurii Maravin, and
  Bharat Ratra.
\newblock {Gravitational Radiation from Primordial Helical Inverse Cascade MHD
  Turbulence}.
\newblock {\em Phys. Rev. D}, Vol.~78, p. 123006, 2008.
\newblock [Erratum: Phys.Rev.D 79, 109901 (2009)].

\bibitem{Kahniashvili:2009mf}
Tina Kahniashvili, Leonard Kisslinger, and Trevor Stevens.
\newblock {Gravitational Radiation Generated by Magnetic Fields in Cosmological
  Phase Transitions}.
\newblock {\em Phys. Rev. D}, Vol.~81, p. 023004, 2010.

\bibitem{Caprini:2009yp}
Chiara Caprini, Ruth Durrer, and Geraldine Servant.
\newblock {The stochastic gravitational wave background from turbulence and
  magnetic fields generated by a first-order phase transition}.
\newblock {\em JCAP}, Vol.~12, p. 024, 2009.

\bibitem{Binetruy:2012ze}
Pierre Binetruy, Alejandro Bohe, Chiara Caprini, and Jean-Francois Dufaux.
\newblock {Cosmological Backgrounds of Gravitational Waves and eLISA/NGO: Phase
  Transitions, Cosmic Strings and Other Sources}.
\newblock {\em JCAP}, Vol.~06, p. 027, 2012.

\bibitem{Giese:2020rtr}
Felix Giese, Thomas Konstandin, and Jorinde van~de Vis.
\newblock {Model-independent energy budget of cosmological first-order phase
  transitions\textemdash{}A sound argument to go beyond the bag model}.
\newblock {\em JCAP}, Vol.~07, No.~07, p. 057, 2020.

\bibitem{Giese:2020znk}
Felix Giese, Thomas Konstandin, Kai Schmitz, and Jorinde van~de Vis.
\newblock {Model-independent energy budget for LISA}.
\newblock {\em JCAP}, Vol.~01, p. 072, 2021.

\bibitem{Espinosa:2010hh}
Jose~R. Espinosa, Thomas Konstandin, Jose~M. No, and Geraldine Servant.
\newblock {Energy Budget of Cosmological First-order Phase Transitions}.
\newblock {\em JCAP}, Vol.~06, p. 028, 2010.

\bibitem{Wainwright:2011kj}
Carroll~L. Wainwright.
\newblock {CosmoTransitions: Computing Cosmological Phase Transition
  Temperatures and Bubble Profiles with Multiple Fields}.
\newblock {\em Comput. Phys. Commun.}, Vol. 183, pp. 2006--2013, 2012.

\bibitem{Guada:2020xnz}
Victor Guada, Miha Nemev\v{s}ek, and Matev\v{z} Pintar.
\newblock {FindBounce: Package for multi-field bounce actions}.
\newblock {\em Comput. Phys. Commun.}, Vol. 256, p. 107480, 2020.

\bibitem{Corbin:2005ny}
Vincent Corbin and Neil~J. Cornish.
\newblock {Detecting the cosmic gravitational wave background with the big bang
  observer}.
\newblock {\em Class. Quant. Grav.}, Vol.~23, pp. 2435--2446, 2006.

\bibitem{Kudoh:2005as}
Hideaki Kudoh, Atsushi Taruya, Takashi Hiramatsu, and Yoshiaki Himemoto.
\newblock {Detecting a gravitational-wave background with next-generation space
  interferometers}.
\newblock {\em Phys. Rev. D}, Vol.~73, p. 064006, 2006.

\end{thebibliography}

\end{document}